\documentclass[twocolumn,english,superscriptaddress,amsmath,floatfix,longbibliography,floats,prb]{revtex4-2}

\usepackage[colorlinks=true,urlcolor=blue,citecolor=blue,linkcolor=blue]{hyperref} 
\usepackage[T1]{fontenc}
\usepackage[utf8]{inputenc}
\usepackage{amssymb}
\usepackage{graphicx}
\usepackage{amsmath,color}
\usepackage{mathrsfs}
\usepackage{float}
\usepackage{indentfirst}
\usepackage{txfonts}
\usepackage[normalem]{ulem}
\usepackage[table]{xcolor}
\usepackage{array,ragged2e}
\usepackage{grffile}
\usepackage{multirow}
\usepackage{algpseudocode}
\usepackage{epstopdf}
\usepackage{booktabs}
\makeatletter

\usepackage{amssymb}
\usepackage{graphicx}
\usepackage{xcolor}
\usepackage{dcolumn}
\usepackage{bm}
\usepackage{dsfont}
\usepackage{appendix}
\usepackage{placeins}

\begin{document}

\title{Floquet dynamical chiral spin liquid at finite frequency}

\author{Didier Poilblanc}
\affiliation{Laboratoire de Physique Th\'eorique, C.N.R.S. and Universit\'e de Toulouse, 31062 Toulouse, France}

\author{Matthieu Mambrini}
\affiliation{Laboratoire de Physique Th\'eorique, C.N.R.S. and Universit\'e de Toulouse, 31062 Toulouse, France}

\author{Nathan Goldman}
\affiliation{Laboratoire Kastler Brossel, Coll\`ege de France, CNRS, ENS-Universit\'e PSL, Sorbonne Universit\'e, 11 Place Marcelin Berthelot, 75005 Paris, France}
\affiliation{Center for Nonlinear Phenomena and Complex Systems, Universit\'e Libre de Bruxelles, CP 231, Campus Plaine, B-1050 Brussels, Belgium}

\date{\today}


\begin{abstract}
    
Chiral Spin Liquids (CSL) are quantum spin analogs of electronic Fractional Chern Insulators. Their realizations on ultracold-atom or Rydberg-atom platforms remain very challenging. Recently, a setup of time-periodic modulations of nearest-neighbor Heisenberg couplings applied on an initial genuine spin liquid state on the square lattice has been proposed to stabilize a (Abelian) $\mathbb{Z}_2$ CSL phase. In the high-frequency limit, it was shown that time evolution can be described in terms of a static effective chiral Hamiltonian. Here we revisit this proposal and consider drives at lower frequency in a regime where the high-frequency Magnus expansion fails. We show that a {\it Dynamical CSL} (DCSL) is nevertheless stabilized in a finite range of frequency. 
The topological nature of this dynamical phase, as well as its instability below a critical frequency, is connected to specific features of the Floquet pseudo-energy spectrum. We also show that the DCSL can be represented faithfully by a two-dimensional time-periodic tensor network and, as in the static case, topological order is associated to a tensor gauge symmetry ($\mathbb{Z}_2$ in that case).

\end{abstract}

\maketitle

\section{Introduction}

Solving correlated models of condensed matter physics is a notoriously hard task for classical computers. 
Quantum simulators based on cold atoms trapped in two-dimensional (2D) optical lattices offer new realistic perspectives in emulating correlated electronic systems~\cite{gross2017quantum,bohrdt_review,daley2022practical}. Currently, new experimental tools and theoretical concepts are being developed for creating topologically-ordered states using ultracold atoms in optical lattices~\cite{Goldman2016,Cooper_review}, including methods based on periodic driving~\cite{Eckardt_review,weitenberg2021tailoring}. Optical-lattice settings were recently used to create a Laughlin state of bosonic atoms~\cite{leonard2023realization}, as well as the symmetry-protected Haldane phase in Fermi–Hubbard ladders~\cite{Sompet2022}; see also Ref.~\cite{lunt2024realization} on the realization of a Laughlin state of rotating fermionic atoms without a lattice.
Rydberg-atom platforms can also be used to produce 2D quantum spin models with tunable interactions~\cite{Ebadi2021,Shai_Floquet}, realizing quantum spin liquids~\cite{Giudici2022} and symmetry-protected topological phases of interacting bosons~\cite{Browaeys_SPT}. 
The toric code, a quantum spin system with ($\mathbb{Z}_2$) topological order, has been addressed on such a Rydberg atom device~\cite{Semeghini2021} and 
on a superconducting quantum processor~\cite{Satzinger2021}.
A wavefunction with more involved topological order has also been produced on a trapped-ion 
quantum processor, allowing for the braiding of non-Abelian anyons~\cite{Iqbal2024}.

Chiral Spin Liquids (CSL) with topological order are also of great interest.
A (non-Abelian) CSL is stabilized in the Kitaev honeycomb model in the presence of a chiral term~\cite{Kitaev2006} and various platforms to implement such a phase have been proposed theoretically using cold atoms~\cite{Sun2023} or Rydberg simulators~\cite{Kalinowski2023}. Both schemes rely on Floquet engineering, i.e. on the action of a periodic drive. 

SU(2)-invariant CSL are also known to exist in spin-1/2 Heisenberg models with a (spin-rotation symmetric) T-breaking term on kagome~\cite{Bauer2014,Niu2022} or
square~\cite{Poilblanc2017b} lattices. However,
designing atomic platforms to prepare SU(2)-symmetric spin-1/2 CSL with full spin-rotational symmetry is still an open challenge. 
Proposals based on synthetic gauge fields to emulate M-colors SU($M$) Hubbard models on the square lattice with 1 fermion per site were shown to be effective for $M\ge 3$~\cite{Chen2016}. Besides, a Floquet-engineering protocol to prepare a SU(2)-symmetric spin-1/2 CSL was proposed in Ref.~\cite{Mambrini2024} by two of the authors.

Here, we shall investigate the applicability of such a Floquet-engineering protocol in the regime where the high-frequency Magnus expansion~\cite{Magnus1954} of the Floquet Hamiltonian~\cite{Floquet1883} may not converge (or would converge very slowly so that many orders of the expansion would be required). We shall consider the setup of Ref.~\cite{Mambrini2024} designed to prepare a topological CSL, starting from a genuine spin liquid state. The high-frequency limit of the protocol (see below) can be entirely described in terms of a static and local Floquet Hamiltonian, breaking time reversal and parity symmetries and hosting a CSL phase. Nevertheless it is not clear whether such a topological phase survives away from the infinite-frequency limit, where the mapping to a {\it static, local and uniform} chiral Floquet Hamiltonian breaks down. We therefore revisit this problem at intermediate frequency on a finite (periodic) system. We find a range of frequency where a Dynamical CSL (DCSL) is still stable, and a critical frequency below which we predict a quantum chaotic behavior. We characterize these different regimes from various properties of the time-evolving state, like its emerging plaquette chirality, using an exact (up to negligible Trotter errors) numerical evaluation on a finite torus. The time-periodic DCSL, stabilized in some stationary regime, is further characterized from its direct relation to a particular eigenstate at the edge of the Floquet (many-body) spectrum. Its topological nature is inferred thanks to a precise tensor network representation in terms of a chiral time-periodic Projected Entangled Pair State. Similarities and differences with respect to the static CSL (obtained in the infinite drive frequency limit) are outlined.

\section{Preparation of the Dynamical Chiral Spin Liquid (DCSL)}
\subsection{Time-periodic drive and symmetries}
\label{subsection:drive}

We start from a spin-1/2 Heisenberg Hamiltonian $H_0=\sum_i J_1 ({\bf S}_i\cdot {\bf S}_{i+e_x} + {\bf S}_i\cdot {\bf S}_{i+e_y})$ on a square lattice, $J_1>0$ being the nearest-neighbor (NN) SU(2)-symmetric antiferromagnetic coupling and $e_x$, $e_y$ the unit vectors along the crystal axis.
The time-dependent Hamiltonian is constructed by adding a periodic drive,  $H(t)=H_0 + H_{\rm drive}(t)$. We consider the time evolution of a quantum state $|\Psi(t)\big> = U(t;0) |\Psi_0\big>$, starting from an initial state $|\Psi_0\big>$ and 
\begin{equation}
    U(t;t_0)={\cal T}_{t'}\exp{(-i\int_{t_0}^{t} H(t') dt')}
    \label{eq:Floquet_unit}
\end{equation}
is the time evolution operator from time $t_0$ to time $t$, involving time-ordering ${\cal T}_{t'}$. Throughout  $\hslash$ is set to 1.

The periodic sequence considered in this work is an adaptation of the driving protocol introduced in Ref.~\cite{Rudner2013} to the many-body quantum spin context~\cite{Mambrini2024}. The original setting of Ref.~\cite{Rudner2013} involves a tight-binding model of non-interacting particles, defined on the square lattice, in which hopping amplitudes are varied in a time-periodic manner. Here, we consider a sinusoidal modulation of the Heisenberg spin-spin interactions dephased by $2\pi/4$ on four types of NN bonds~\cite{Mambrini2024}
\begin{eqnarray}
    H_{\rm drive}(t)&=&\cos{(\omega t)}H_x+\sin{(\omega t)}H_y, \\
    {\rm where} \hskip 0.3cm
    H_x&=&J\sum_{i\in A} ({\bf S}_i\cdot {\bf S}_{i+e_x} - {\bf S}_i\cdot {\bf S}_{i-e_x}) \nonumber\\
    {\rm and} \hskip 0.3cm H_y&=&J\sum_{i\in A} ({\bf S}_i\cdot {\bf S}_{i+e_y} - {\bf S}_i\cdot {\bf S}_{i-e_y}).\nonumber
    \end{eqnarray}
where $J$ (set to $1$ unless specified) is the coupling constant, and where the site summation is performed over the sites $i\!=\!(i_x,i_y)$ belonging to one of the two sublattices ($A$) of the square lattice, $i_x+i_y$ even, to avoid double-counting of the bonds. 
$H_{\rm drive}(t)$ has been set up to be chiral, i.e. breaking parity $P$ (reflection w.r.t. a crystal axis) and time-reversal $T$ ($t\rightarrow -t$) but not the product PT.
The bond pattern is staggered leading to two types of square plaquettes. However, the Hamiltonian $H_0+H_{\rm drive}$ is invariant under the combination of the unit translations with the spatial inversion $I=R_\pi$ ($\pi$-rotation w.r.t. a lattice site) i.e. ${\tilde T_x}=T_x I$ and ${\tilde T_y}=T_y I$.
The (non-commuting) symmetries ${\tilde T_x}$ and ${\tilde T_y}$ play a key role as will be discussed later on. 
The Hamiltonian $H$ is also invariant under SU($2$) spin rotations. Hence, time evolution starting from a singlet wavefunction will occur in the spin-singlet manifold.

\subsection{Floquet Hamiltonian and heating}

Time evolution at stroboscopic times, i.e. at time intervals multiple of the drive period $T_{\rm drive}$, is provided by the unitary operator, 
\begin{equation}
    U(t_0+T_{\rm drive};t_0)=\exp{(-iH_F[t_0] T_{\rm drive})}\, .
    \label{eq:Floquet_unit2}
\end{equation}
The above equation defines the (hermitian) Floquet operator $H_F$, which characterizes the time-evolution over one driving period. Here, we have explicitly kept a time degrees of freedom $t_0$: 
although the spectrum of $H_F[t_0]$ is generically independent of $t_0$ (indeed,  
$
    H_F[t_0]=R(t_0)^\dagger H_{{\rm eff}} R(t_0)
$, 
where $R(t_0)$ is a unitary matrix and where $H_{{\rm eff}}$ is a time-independent effective Hamiltonian~\cite{Goldman_Dalibard,Rahav2003,Bukov2015}) the dependence of the Floquet eigenstates on the time parameter $t_0$ will be of particular interest. 
From Eq.~(\ref{eq:Floquet_unit2}) it is easy to see that the "quasi-energy" spectrum $\{E\}$ of $H_F$ is defined modulus $2\pi/T_{\rm drive}$, i.e in the Floquet-Brillouin zone (FBZ), $E\in [-\omega/2,\omega/2]$. 
Hereafter, we shall implicitly assume that our system is finite, constituted of $N$ quantum spins (or qubits). For a many-body system we expect the width of the energy spectrum (named "bandwidth" hereafter) to scale like the volume, i.e. $W_N\propto N$, and to eventually cover the full FBZ when increasing size. In that case we expect the system to heat up. 

One can obtain a more precise estimation of the bandwidth from the $\omega\gg J$ limit studied previously~\cite{Mambrini2024}. In that case the (leading-order of the) effective Floquet Hamiltonian can be obtained~\cite{Goldman_Dalibard,Bukov2015}, e.g. for our Hamiltonian~\cite{Mambrini2024},
\begin{equation}
    H_F^0 =H_0+ \frac{i}{2\omega} [H_x,H_y]
    =H_0 + J_F \, i\sum_\square (P_{ijkl} - P_{ijkl}^{-1}), 
    \label{eq:FloquetHam}
\end{equation}
where $J_F=\frac{J^2}{4\omega}$ and $P_{ijkl}$ is the 4-site cyclic permutation which applies to all plaquettes $\square=(ijkl)$. 
Note that the (imaginary part of the) cyclic permutation on the  (oriented) plaquette $(i,j,k,l)$  has a simple expression in terms of spin operators,
\begin{eqnarray}
i (P_{ijkl} - P_{ijkl}^{-1})
 &=& 2\, \{ \, {\bf S}_i\cdot ({\bf S}_{j}\times {\bf S}_{l})
+{\bf S}_{j}\cdot ({\bf S}_{k}\times {\bf S}_{i})\nonumber \\
    && +\,{\bf S}_{k}\cdot ({\bf S}_{l}\times {\bf S}_{j})
    +{\bf S}_{l}\cdot ({\bf S}_{i}\times {\bf S}_{k})\, \}  ,     
 \label{eq:cyclic}
   \end{eqnarray}
   being the sum of the (oriented) chiralities on the four triangles of the plaquette. We point out that the Floquet Hamiltonian $H_F[t_0]\!=\!H_F^0$ in Eq.~\eqref{eq:FloquetHam} does not depend on the time parameter $t_0$ in the high-frequency limit:~indeed, using the high-frequency expansion of Refs.~\cite{Goldman_Dalibard,Bukov2015}, we find that the leading $t_0$-dependent term is of the order of $JJ_1/\omega\sim J^3/\omega^2$, and hence, can be neglected. In this sense, any substantial dependence of the Floquet eigenstates on the time parameter $t_0$ will signal the onset of the low-frequency regime.
Qualitatively, assuming $J_1\sim J_F$ (see later on), we expect the bandwidth to behave as 
\begin{equation}
    W_N(\omega) = aNJ_F=aNJ^2/4\omega\, ,
    \label{bandwidth}
\end{equation}
where $a$ is a numerical factor, $a\simeq 2.4$ in the high-frequency limit. 
Since the Floquet energies are defined modulus $\omega$, one can define a "gap" in the Floquet-Brillouin energy zone, 
\begin{equation}
    \Delta_N(\omega)=\omega-W_N(\omega)\, ,
    \label{FBgap}
\end{equation}
and the vanishing of the gap (obtained by increasing $N$ or decreasing $\omega$) may lead to a chaotic behavior. 
Assuming $a$ weakly depends on $\omega$ as confirmed numerically (see Appendix~\ref{eq:Floquet_unit} for the exact calculation of the Floquet quasi-energy spectrum), this simple analysis provides an estimate of the critical frequency, 
\begin{equation}
    \omega_c/J\simeq \frac{1}{2}(aN)^{1/2}\, , 
    \label{crit_freq}
\end{equation}
below which heating quickly occurs. In the last part of this article we shall consider a $4\times 4$ torus of $N=16$ qubits, leading to $\omega_c\simeq 3.1$. This crude argument turns out to be qualitatively correct as shown below.

\subsection{Adiabatic evolution}

Following the quasi-adiabatic protocol introduced in Ref.~(\cite{Mambrini2024}), we consider the possibility of ramping the amplitude $J$ of the drive by some time-dependent factor $\lambda(t)$ slowly increasing between $0$ and $1$. 
\begin{equation}
      H(t)= H_0 + \lambda(t) H_{\rm drive}(t)\, .
      \label{eq:Htotal}
\end{equation}
Here we choose
\begin{equation}
    \lambda(t)=(1-\cos{(\pi t / t_{\rm ramp})})/2\, ,
\end{equation}
where $t_{\rm ramp}=n_{\rm ramp} T_{\rm drive}$ is the ramp time. The ramp function is chosen so that the "ramp velocity" $\partial \lambda/\partial  t$ vanishes at the initial and final times, $t=0$ and $t=t_{\rm ramp}$, to improve the adiabaticity of the path. 

\begin{figure}
	\centering
	\includegraphics[width=0.97\columnwidth]{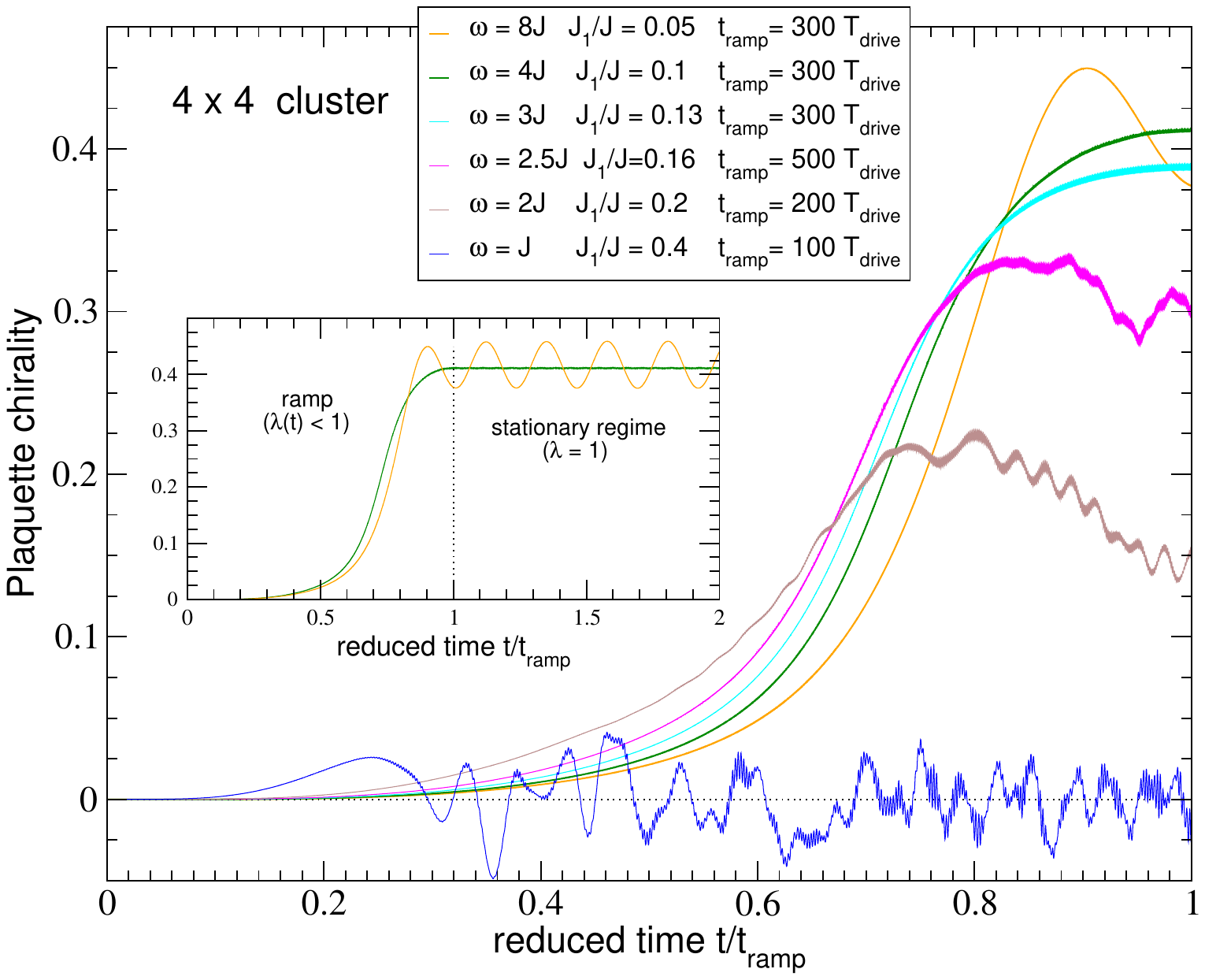}
	\caption{Expectation value of the plaquette chirality ${\rm Im}\big< P_{ijkl}\big>$ versus time under a slow ramping of the drive at various frequencies. Time is shown in reduced units $t/t_{\rm ramp}$. A stationary regime is obtained for $\omega\ge 3$ (see inset for $\omega=4$ and $8$) while a chaotic behavior is observed for $\omega\le 2.5$. The critical value $\omega_c$ is expected to lie between 2.5 and 3.}
	\label{fig:ramp}
\end{figure}

Note that the antiferromagnetic Heisenberg coupling $J_1$ of $H_0$ is also an essential parameter in our setup. In the high-frequency limit, $J_1$ has to be adjusted to be of the same order as $J_F\propto 1/\omega$ so that one sits in the region of stability of the CSL of $H_F^0$. Here we assume a similar relation holds for all frequency, choosing $J_1=0.4/\omega$. 
We then expect the final state at $t=t_{\rm ramp}$ to be in a CSL phase. 

We have tested this scenario numerically on a $4\times 4$ torus, starting from the (singlet, translation and $C_{4v}$-invariant) ground state $|\Psi_0\big>$ of $H_0$ and applying the unitary time-evolution $U(t;0)={\cal T}_{t'}\exp{(-i\int_{0}^{t} H(t') dt')}$ to it. A fine Trotter discretization of U(t) is used~\cite{Trotter1959,SUZUKI1990} -- typically $N_\tau=64$ Trotter steps are used within each time period -- and the calculation is performed in the $S_z=0$ basis (${\it dim}=12\, 870$). The expectation value of the (uniform) plaquette chirality ${\rm Im} \big< P_{ijkl}\big>$ in the time-evolved state $|\Psi(t)\big>=U(t)|\psi_0\big>$ is computed as a function of time and results are shown in Fig.~\ref{fig:ramp} for several frequencies. For intermediate, but large enough frequencies, we observe a smooth increase of this observable with a saturation when $t\rightarrow t_{\rm ramp}$ and a plateau for $t>t_{\rm ramp}$ ($\lambda(t)$ is set to 1 from there). This is the first sign that a CSL phase is stabilized in this regime, to be corroborated by further results later on. In contrast, at too low frequency $\omega <\omega_c$ no plateau is reached, with a first smooth rise transiting eventually to a chaotic behavior as e.g. for $\omega=2.5$, or showing almost immediately a direct chaotic behavior as e.g. at $\omega=1$. The change of behavior at a frequency around $\omega_c\sim 2.7$ is consistent with the analysis of the Floquet quasi-energy spectrum reported in Appendix~\ref{app:Floquet} and the crude estimation in Eq.~\eqref{crit_freq}; heating occurs when the Floquet bandwidth $W_N(\omega)$ reaches $\omega$, leading to a chaotic behavior in the time evolution. 

\begin{figure}
	\centering
	\includegraphics[width=0.97\columnwidth]{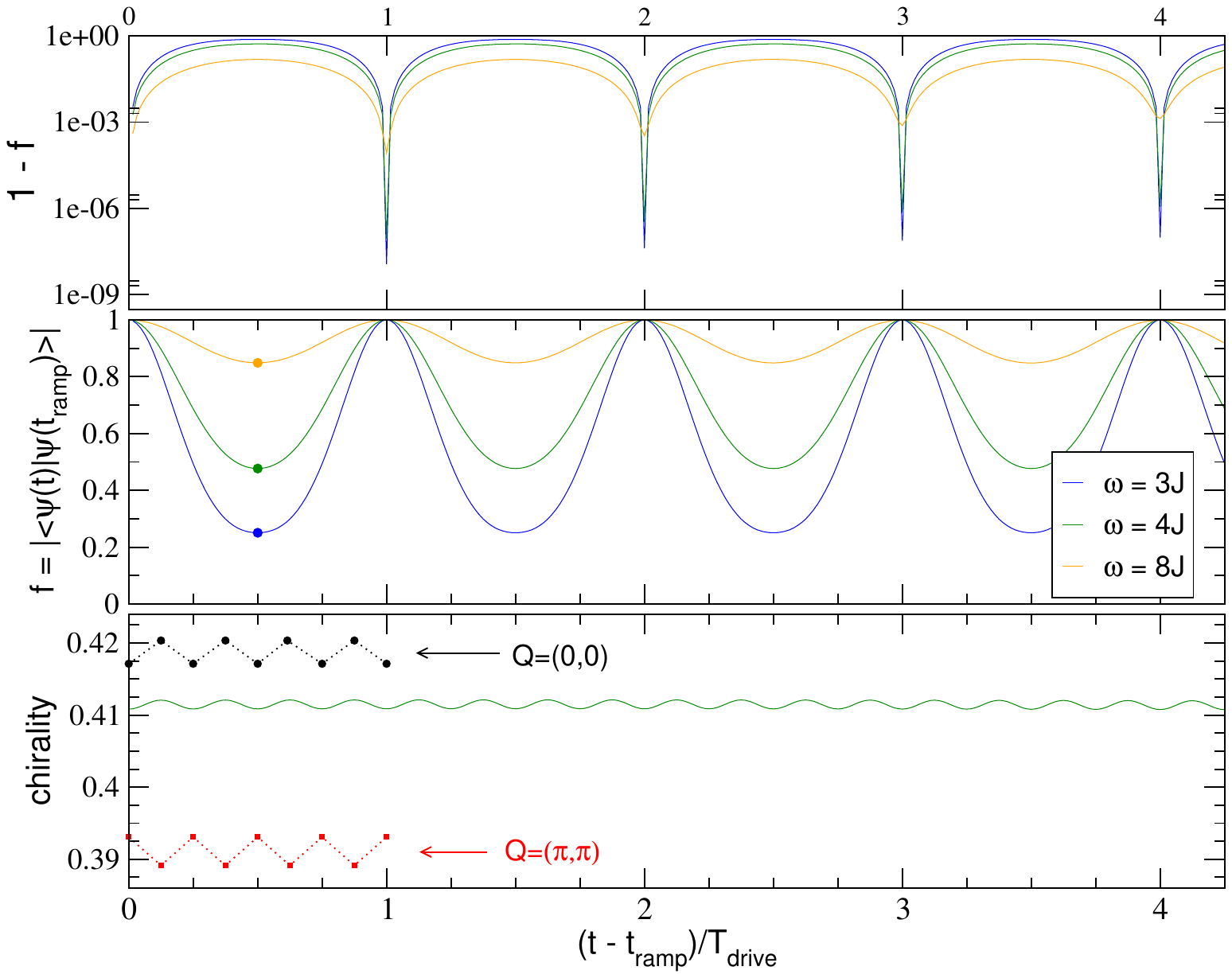}
	\caption{Stationary regime $t\ge t_{\rm ramp}$. Top panels: $f=|\big<\Psi(t)|\Psi(t_{\rm ramp}\big>|$ and $1-f$ as a function of time, showing Rabi oscillations. Bottom panel: plaquette chiralities $\frac{1}{N} \big<H_\square\big>$ in the state $|\Psi(t)\big>$ and in its two components $|\Psi_{0}^+\big>$ and $|\Psi_{\pi}^-\big>$ of momenta $Q=(0,0)$ and $Q=(\pi,\pi)$, respectively. A smaller periodicity $\frac{1}{4}T_{\rm drive}$ of the plaquette chiralities is visible. Note the dots in the middle panel at $t_0=t-t_{\rm ramp} = \frac{1}{2}T_{\rm drive}$ computed from $f=a_0-a_1=2a_0-1$, where the weights $a_0$ and $a_1$ are defined in the text and shown in Fig.~\ref{fig:finiteQ}.
 }
	\label{fig:stationnary}
\end{figure}

\section{Stationary regime}

We now concentrate on the stationary regime for $\omega>\omega_c$ and $t>t_{\rm ramp}$. Our goal is to establish, step by step, the precise form of the periodic time-dependent state $|\Psi(t)\big>$.

\subsection{Rabi oscillations}

Figure~\ref{fig:stationnary} shows in the center panel the fidelity
$f=|\big<\Psi(t)|\Psi(t_{\rm ramp})\big>|$, taking as reference state the state at $t=t_{\rm ramp}$. One sees clear Rabi oscillations. Indeed, as shown in the top panel, the state $|\Psi(t_{\rm ramp}\big>$ returns to itself periodically, with a very high accuracy.  This suggests the presence of two components in $|\Psi(t)\big>$. Assuming continuity with the large-$\omega$ limit, one component $|\Psi^{+}_{0}\big>$ should be invariant under lattice translation and under spatial inversion $I=R_\pi$, which are the (independent) symmetry properties of the $\omega\rightarrow\infty$ ground state of the Floquet Hamiltonian~(\ref{eq:FloquetHam}). At finite $\omega$, the Hamiltonian $H(t)$ is only invariant under the {\it products} ${\tilde T_x}=T_x I$ and ${\tilde T_y}=T_y I$, but {\it not} under the elementary lattice translations and the inversion separately. Therefore, $|\Psi^{+}_{0}\big>$ state is coupled under time evolution to a component $|\Psi^{-}_{\pi}\big>$ which is odd under inversion and unit translations (momentum $(\pi,\pi)$). For any arbitrary time $t_{\rm ramp} + {t_0}$ after the ramp we can write, 
\begin{equation}
    |\Psi[t_0]\big> \equiv |\Psi(t_{\rm ramp} + {t_0})\big> = |\Psi^{+}_{0}[t_0]\big> + \exp{(i\omega t_0)}\, |\Psi^{-}_{\pi}[t_0]\big>\, .
    \label{eq:psi}
    \end{equation}
This formula is fully consistent with the Rabi oscillations observed in the upper panel of Fig.~\ref{fig:stationnary} provided the two components $|\Psi^{+}_{0}[t_0]\big>$ and $|\Psi^{-}_{\pi}[t_0]\big>$ are time-independent or, at least, weakly time dependent (as it will become clear later on). 
In fact, a closer look reveals small oscillations of period $\frac{1}{4}T_{\rm drive}$ of the chirality on triangles, as shown in the bottom panel of Fig.~\ref{fig:stationnary}, and which can be attributed to small variations of $|\Psi^{+}_{0}[t_0]\big>$ and $|\Psi^{-}_{\pi}[t_0]\big>$ as a function of time. Interestingly, we observe that, periodically at "high-symmetry times" $t_0^{(n)}=n\frac{1}{4} T_{\rm drive}$, the chirality on triangles computed separately in $|\Psi^{+}_{0}[t_0^{(n)}]\big>$ or in $|\Psi^{-}_{\pi}[t_0^{(n)}]\big>$ bears the full $C_{4v}$ point group symmetry of the square lattice (in particular, are invariant under $\pi/2$-rotations). In fact, this emerging $\frac{1}{4}T_{\rm drive}$ periodicity applies only to specific observables. Generically, spatial anisotropy still occurs, e.g. a small (time-dependent) difference between the spin-spin correlations along the $x$ or $y$ directions is seen, even at times $t_0^{(n)}$. We will show later-on that 
the two components $|\Psi^{+}_{0}[t_0]\big>$ and $|\Psi^{-}_{\pi}[t_0]\big>$ are themselves linear combination (with time dependent amplitudes) of two states belonging to the one-dimensional irreducible representations (IRREPs) ${A}$ and ${B}$ on one hand, and ${E}$ and ${E'}$ on the other hand, as defined in Table~\ref{tab:sym}. 

\begin{figure}
	\centering
	\includegraphics[width=0.97\columnwidth]{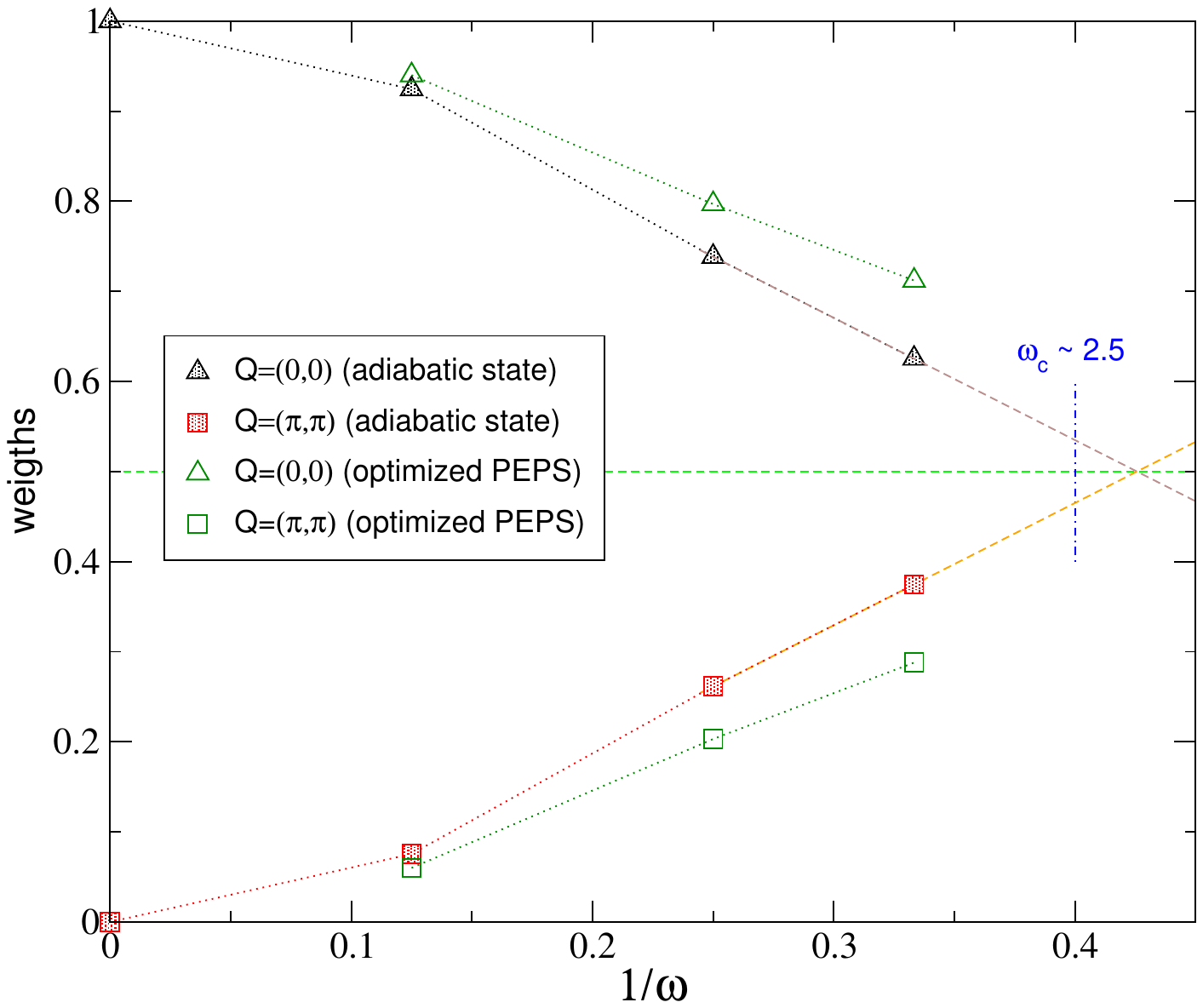}
	\caption{
Stationary regime: weights $a_0$ and $a_\pi$ of the two components of the stationary state appearing in (\ref{eq:psi}) (shown as filled symbols) versus inverse-frequency. For comparison the weights of the decomposition of the optimized PEPS $|\Psi_{\rm PEPS}^{0}\big>$ are also shown (open symbols).
	}
	\label{fig:finiteQ}
\end{figure}

In the high-frequency limit, the Floquet Hamiltonian (\ref{eq:FloquetHam}) has a higher symmetry being invariant independently under the units translations and the inversion (and not only a combination of those). As shown in a previous work, its GS also has higher symmetry and one of the two components of (\ref{eq:psi}) should vanish in that limit. We have tested this prediction by computing the weights $a_0=|\big<\Psi[t_0]|\Psi^{+}_{0}[t_0]\big>|^2$ and
$a_\pi=|\big<\Psi[t_0]|\Psi^{-}_{\pi}[t_0]\big>|^2=1-a_0$. The $Q=0$ and $Q=(\pi,\pi)$ components of the wavefunction are obtained as $\frac{1}{2}\{|\Psi[t_0]\big> \pm T_{x}|\Psi[t_0]\big>\}$. The above weights have been computed at various times $t_0^{(n)}=\frac{n}{4} T_{\rm drive}$, showing no dependence on $t_0$, and results are shown in Fig.~\ref{fig:finiteQ}. As expected, $a_0$ ($a_1$) which is unity (zero) in the high-frequency limit $1/\omega=0$, steadily decreases (increases) down (up) to $\sim 0.6$ ($\sim 0.4$) at $\omega=3$, before the transition. Note the proximity between the (estimated) transition point $\omega_c\sim 2.7$ and crossing point $\omega_\times\sim 2.5$ between $a_0$ and $a_1$.

\subsection{Floquet eigenstates}

To get a complete understanding of the structure of the stationary wavefunction and of the breakdown of the stationary regime, it is helpful to investigate  the Floquet quasi-energy spectrum in more details. The latter is defined within a Floquet-Brillouin zone, which we take to be the energy range $[-\omega/2,\omega/2]$, i.e. the natural Floquet zone. Details about its derivation and results can be found in Appendix~\ref{app:Floquet}. 
As previously anticipated, we indeed find that the spectrum bandwidth increases like $1/\omega$ upon reducing the frequency~[Eq.~\eqref{bandwidth}]. We estimate numerically (see Fig.~\ref{fig:spectrum2}) the critical frequency at which the many-body bandwidth $W_N(\omega)$ reaches the size of the Floquet-Brillouin zone to be around $\omega_c\simeq 2.7$. For $\omega\le \omega_c$ heating is expected to occur: in that case the quasi-adiabatic protocol breaks down, and the stationary plateau is replaced by a chaotic behavior, in agreement with our previous findings shown in Fig~\ref{fig:ramp}. Note that heating may also occur for $\omega> \omega_c$ but this is expected only after an exponentially long time with respect to the natural time-scale $\tau_E=1/\Delta_N(\omega)=1/(\omega-W_N(\omega))$~\cite{Fleckenstein2021}. 

\begin{table}
    \centering
    \begin{tabular}{lcccccccc} \toprule   
    Symmetries & $T_x R_\pi$ & $T_y R_\pi$ & $R_{\pi/2}$ & $R_\pi$& $R_x \Theta$ & $R_{x+y}\Theta$ & $R_{y}\Theta$ & $R_{x-y}\Theta$\\ 
    \midrule
       $Q=(0,0)$ / $A$ IRREP  & +1 & +1 & +1& +1 &+1& +1 & +1 & +1  \\
      $Q=(0,0)$ / $B$ IRREP  & +1 & +1 & $-1$& +1 &+1& $-1$ & +1 & $-1$  \\
       $Q=(\pi,\pi)$ / $E$ IRREP & $+1$ & $+1$ & $+i$ & $-1$&+1& $+i$ & $-1$ & $-i$ \\   
       $Q=(\pi,\pi)$ / $E'$ IRREP & $+1$ & $+1$ & $-i$ & $-1$ &$-1$& $+i$ & $+1$ & $-i$ \\    
       \bottomrule
    \end{tabular}
    \caption{Chiral space group: characters ($\pm 1$, $\pm i$) under some relevant group operations (listed on the first line) of the four components of the stationary state. As the Floquet Hamiltonian all components are invariant under unit translations combined with spatial inversion ($\pi$-rotation w.r.t. a lattice site) -- see second and third columns. Axis reflections $R_x$, $R_y$, $R_{x+y}$ and $R_{x-y}$ are combined with complex conjugation $\Theta$ (time-reversal). Hence the characters associated to the (combined) reflections are defined up to a global phase. 
    }
    \label{tab:sym}
\end{table}

The dependence of $H_F[t_0]$ and of the Floquet eigenstates $|\phi_\alpha[t_0]\big>$ (labeled by $\alpha\in\mathbb{Z}$, see later) on the time parameter $t_0$ is of particular interest. 
First,  restricting here to eigenstates combining momenta $q=(0,0)$ and $q=(\pi,\pi)$, one can write:
\begin{equation}
    |\phi_\alpha[t_0]\big> = |\phi^{+}_{0,\alpha}[t_0]\big> + \exp{(i\omega t_0)}\, |\phi^{-}_{\pi,\alpha}[t_0]\big>\, ,
    \label{eq:phi}
    \end{equation}
 where the two components are even and odd under spatial inversion, respectively. 
Secondly, we note that a shift $t_0\rightarrow t_0+T_{\rm drive}/4$ corresponds to rotate the time-dependent bonds by $90$-degree, 
$H(t+\frac{1}{4}T_{\rm drive})=R_{\pi/2}H(t)R_{\pi/2}^\dagger$, 
which is physically equivalent. Then, one can write $\exp{(-iH_F[t_0 +\frac{1}{4}T_{\rm drive}] T_{\rm drive})}$ as:
\begin{eqnarray}
&&{\cal T}_{t'}\exp{(-i\int_{t_0+\frac{1}{4}T_{\rm drive}}^{t_0+\frac{5}{4}T_{\rm drive}} H(t') \, dt')}\nonumber\\
    &&={\cal T}_{t'}\exp{(-i\int_{t_0}^{t_0+T_{\rm drive}} H(t'+\frac{1}{4}T_{\rm drive})\, dt')}\nonumber\\
              &&={\cal T}_{t'}\exp{(-i R_{\pi/2}\int_{t_0}^{t_0+T_{\rm drive}} H(t')\, dt'\, R_{\pi/2}^\dagger) }\, ,
        \label{eq:Floquet_period}
\end{eqnarray}
which shows that
\begin{equation}
    H_F[t_0+\frac{1}{4}T_{\rm drive}]= R_{\pi/2} H_F[t_0] R_{\pi/2}^\dagger\, .
    \label{eq:Floquet_rot}
\end{equation}
Using (\ref{eq:Floquet_rot}) we expect that $|\phi_\alpha[t_0+\frac{1}{4}T_{\rm drive}]\big>=R_{\pi/2} |\phi_\alpha[t_0]\big>$, providing an explicit correspondence between "time" and "space". 
Then, one can propose the following ansatze to ensure the above property:
\begin{eqnarray}
\label{eq:phi2}
|\phi^{+}_{0,\alpha}[t_0]\big> &=&|\phi^{A}_{0,\alpha}\big> + \exp{(2i\omega t_0)}\, |\phi^{B}_{0,\alpha}\big>\, \\
|\phi^{-}_{\pi,\alpha}[t_0]\big> &=&  |\phi^{E}_{\pi,\alpha}\big> + \exp{(-2i\omega t_0)}\, |\phi^{E'}_{\pi,\alpha}\big>\, ,\nonumber
\end{eqnarray}
where the four (time-independent) wavefunctions labeled by $A$, $B$, $E$ and $E'$ belong to the ${A}$, ${B}$, ${E}$ and ${E'}$ IRREPs listed in Table~\ref{tab:sym}, respectively, and are (exactly) periodic of period $T_{\rm drive}/4$. We have checked numerically that the phase factors are indeed correct. Using (\ref{eq:phi}) we can then obtain the full decomposition of the Floquet eigenstates as
\begin{eqnarray}
\label{eq:phi-decomp}
|\phi_\alpha[t_0]\big> &=& |\phi^{A}_{0,\alpha}\big> + \exp{(2i\omega t_0)}\, |\phi^{B}_{0,\alpha}\big>  \\
 &+& \exp{(i\omega t_0)}\,|\phi^{E}_{\pi,\alpha}\big> + \exp{(-i\omega t_0)}\, |\phi^{E'}_{\pi,\alpha}\big>\, ,\nonumber
 \end{eqnarray}
where the four states entering the decomposition are only very weakly time-dependent but, strictly, time-periodic of period $T_{\rm drive}/4$. In fact, the above classification of the Floquet eigenstates is incomplete. A careful analysis of the symmetry of the Floquet unitary in Appendix~\ref{app:FloquetSym} shows that there is another class of eigenstates 
involving a new set of four states with different quantum numbers.
We expect a decomposition of this second class of states (labelled by a tilde) similar to (\ref{eq:phi-decomp}):
\begin{eqnarray}
\label{eq:phi-decomp2} 
|{\tilde \phi}_\alpha[t_0]\big> &=& |\phi^{+}_{\pi,\alpha}[t_0]\big> + \exp{(i\omega t_0)}\, |\phi^{-}_{0,\alpha}[t_0]\big> \nonumber\\
&=& |\phi^{A}_{\pi,\alpha}\big> + \exp{(2i\omega t_0)}\, |\phi^{B}_{\pi,\alpha}\big> \\
 &+& \exp{(i\omega t_0)}\,|\phi^{E}_{0,\alpha}\big> + \exp{(-i\omega t_0)}\, |\phi^{E'}_{0,\alpha}\big>\, ,\nonumber
\end{eqnarray}
where, in contrast to (\ref{eq:phi-decomp}), the $A$ and $B$ ($E$ and $E'$) IRREPs are associated to momentum $(\pi,\pi)$ ($(0,0)$).
The phase factors have been set again using the correspondence between (discrete) translation in time and (point group) $\pi/2$-rotation. In other words, the phase factors in (\ref{eq:phi-decomp}) and (\ref{eq:phi-decomp2}) have been introduced to capture most of the variation with $t_0$: the eight states, as defined by the equations, are then exactly periodic with period $T_{\rm drive}/4$ and are only very weakly time-dependent within a quarter-period.

\begin{figure}
	\centering
	\includegraphics[width=0.97\columnwidth]{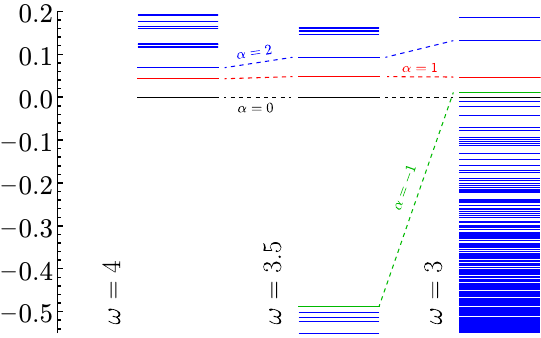}
	\caption{Zoom of the Floquet "low-energy" spectrum around the special eigenstate $\alpha=0$ (taken as the energy reference), in the mixed $Q=0$/$Q=(\pi,\pi)$ sector. Energies on the vertical axis are in units of $\pi/T_{\rm drive}=\omega/2$. A comparison between spectra for $\omega=4$, $\omega=3.5$ and $\omega=3$ is shown, revealing the closing of the gap in the FBZ. The states labeled by $\alpha=0$ and $\alpha=1$ are topological partners. 
  Note the (anti)crossing between the $\alpha=0$ and $\alpha=-1$ states between $\omega=3.5$ and $\omega=3$. }
	\label{fig:floquet_zoom}
\end{figure}

\subsection{Adiabaticity and micromotion}

The time evolution of the quantum state is obtained by a very slow increase of the amplitude of the drive (of constant frequency) so that we expect some adiabatic theorem to apply on such a finite system. In fact, one can continuously follow the eigenvalues of $H_F$ starting from the $\omega\rightarrow\infty$ low-energy states of $H_F^0$ (labeled as $|\phi_\alpha\big>$, $\alpha=0,1,2\cdots$) and decreasing the frequency. At lower frequency, the latter can then still be named as "low-energy" Floquet states, except when the spectrum covers the full quasi-energy FBZ for $\omega<\omega_c$. In fact, when 
$\omega\gtrsim\omega_c$, some eigenvalues approach the $\alpha=0$ eigenvalue from below as shown in Fig.~\ref{fig:floquet_zoom}. We label the corresponding eigenstates as $\alpha=-1,-2,-3,\cdots$, going down away from the $\alpha=0$ eigenstate and qualify them also as "low-energy" eigenstates, for simplicity. 

\begin{table}
    \centering
    \begin{tabular}{lccc} \toprule   
    $\omega$: & 8 & 4 & 3 \\
    \midrule
         $\alpha=3$ & 0 & 0 & 0 \\
    $\alpha=2$& $5.38\,10^{-7}$& $8.43\,10^{-7}$& $3.73\,10^{-7}$ \\  
     $\underline{\alpha=1}$& $1.\,10^{-11}$& $4.\,10^{-10}$ &  $1.3\,10^{-9}$  \\
       \underline{$\alpha=0$}\hspace*{0.3cm} & 0.989\,854\,14 \hspace*{0.3cm} & 0.999\,999\,415\hspace*{0.3cm} & 0.999\,999\,975\\
        $\alpha=-1$ & $3.0\,10^{-8}$ & $4.1\,10^{-9}$ & $1.02\,10^{-8}$ \\
        $\alpha=-2$ & $3.\,10^{-11}$ & $1.\,10^{-11}$ & $2.\,10^{-11}$ \\       
        \\
        $|\big<\Psi_{\rm PEPS}^{0}|\Psi[0]\big>|$ & 0.962\,934  & 0.935\,259 &  0.928\,606\\
        $|\big<\Psi_{\rm PEPS}^{0}|\phi_0[0]\big>|$ & 0.944\,342 & 0.935\,433 &   0.928\,652 \\ 
        $|\big<\Psi_{\rm PEPS}^{1}|\phi_1[0]\big>|$ & 0.882\,121 & 0.788\,390 &  0.825\,449 \\        
        \bottomrule
    \end{tabular}
    \caption{Overlap $f=|\big<\Psi[t_0]|\phi_\alpha[t_0]\big>|$ between the time evolved state (using $t_{\rm ramp}=300 T_{\rm drive}$) in the stationary regime and the Floquet "low-energy" eigenstates. $N_\tau=64$ Trotter steps per drive period are used. We have checked explicitly that overlaps at times $t_0=n\frac{1}{8}T_{\rm drive}$, $n\in\mathbb{N}$, are identical. 
    The overlaps $|\big<\Psi_{\rm PEPS}^{0}|\Psi[0]\big>|$ and $|\big<\Psi_{\rm PEPS}^{0}|\phi_0[0]\big>|$ between the optimized $D=3$ PEPS of Eq.~(\ref{eq:psi-adiab}) and, both, the adiabatic state and the $\alpha=0$ Floquet state (named as the "ground state") are also shown. The last line shows the (normalized) overlap $|(\big<\Psi_{\rm PEPS}^{1}|\phi_1[0]\big>|$ relevant for the other topological sector. All states involved in the overlaps are supposed to be normalized.}
    \label{tab:overlap}
\end{table}

We have computed the overlaps of the time-evolved state $|\Psi(t)\big>$ in the stationary regime (for $t_0=t-t_{\rm ramp}$ multiples of $\frac{1}{8}T_{\rm drive}$) with the low-energy Floquet states $|\phi_\alpha[t_0]\big>$ and found an overlap $f$ very close to 1 with the "ground state" $|\phi_0[t_0]\big>$ and very small overlaps with the other $\alpha\ge 1$ eigenstates,
as shown in Table~\ref{tab:overlap}. Hence, in such an adiabatic process, the stationary time-evolving state $|\Psi[t_0]\big>$ is following closely the lowest energy Floquet state,
\begin{equation}
|\Psi[t_0]\big>\simeq |\phi_0[t_0]\big>\, ,
\label{eq:adiab}
\end{equation}
with a high fidelity (missing weight $1-f^2\sim\!2(1-f)\sim\! 10^{-6}\, /\, 5.\,10^{-8}$ for $\omega=4\, /\, 3$).
Using the (quasi-)adiabaticity given by (\ref{eq:adiab}) and (\ref{eq:phi-decomp}) we obtain the full decomposition of the adiabatic state (up to a global phase),
\begin{eqnarray}
|\Psi[t_0]\big> &\simeq& |\phi^{A}_{0,0}\big> + \exp{(2i\omega t_0)}\, |\phi^{B}_{0,0}\big> \label{eq:psi-adiab} \\
 &+& \exp{(i\omega t_0)}\,|\phi^{E}_{\pi,0}\big> + \exp{(-i\omega t_0)}\, |\phi^{E'}_{\pi,0}\big>\, ,\nonumber
\end{eqnarray}
with very high fidelity, providing a complete expression of the micromotion within a time period. Note that the $B$ ($E'$) state carries only a small fraction of the weight $a_0$ ($a_\pi$) associated to the $Q=0$ ($Q=(\pi,\pi)$) component of the wave function $|\Psi[t_0]\big>$ reported in Fig.~\ref{fig:finiteQ}.

At larger frequency the missing weight is larger, $1-f^2\sim\!2.\,10^{-2}$ at $\omega=8$, suggesting that a slower ramp (i.e. a larger value of $n_{\rm ramp}$) would be needed to achieve the same very high fidelity. Note that an admixture in $|\Psi[0]\big>$ of a (or a few) higher excited state(s) of $H_F$ is consistent with the strong oscillations seen in the inset of Fig.~\ref{fig:ramp}.

\subsection{PEPS representation}

The $\mathbb{Z}_2$ CSL has been shown to be represented faithfully by chiral Projected Entangled Pair States (PEPS)~\cite{Poilblanc2015,Poilblanc2016,Hasik2022}.
It is therefore instructive to find a PEPS representation of our final state prepared at $t=t_{\rm ramp}$. This will provide a more profound characterization of the nature of the state. First, since the time evolution occurs within the singlet manifold, we will restrict ourselves to SU(2)-symmetric PEPS (see Refs.~\cite{Mambrini2016} for details). For simplicity, we will choose the smallest possible virtual space $\nu = 0 \oplus 1/2$ (i.e. a bond dimension $D=3$) which may already provide a faithful representation. In our previous study in the high-frequency limit, we found that the prepared state has $Q=(0,0)$ (the $|\Psi_\pi^-\big>$ component vanishes in that limit) and belongs to the ${A}$ representation of the point group (second line of Table~\ref{tab:sym}). At finite frequency however, one needs to account for the emerging $Q=0$ ${B}$ and $Q=(\pi,\pi)$ $E,E'$ components (see Table~\ref{tab:sym}). For $D=3$ we have found three independent {\it complex} site tensors $A_\alpha$, $B_\beta$ $E_\gamma$ and $E'_\delta$ in each of the four classes defined by the point group characters given in Table~\ref{tab:sym} (see Appendix~\ref{app:peps} for details). The state $|\Psi[0]\big> = |\Psi^{+}_{0}\big> + |\Psi^{-}_{\pi}\big>$ is well captured by a PEPS $|\Psi_{\rm PEPS}^{0}\big>$
constructed from two different but related tensors $T_A$ and $T_B$ on the two $A$ and $B$ sublattices;
\begin{eqnarray}
    T_A &=& \sum_\alpha x_\alpha A_\alpha + \sum_\beta y_\beta B_\beta +\sum_\gamma z_\gamma E_\gamma + \sum_\delta w_\delta {E'}_\delta \nonumber\\
        T_B &=& \sum_\alpha x_\alpha A_\alpha + \sum_\beta y_\beta B_\beta -\sum_\gamma z_\gamma E_\gamma - \sum_\delta w_\delta {E'}_\delta\, , 
\label{eq:peps_tensors}
\end{eqnarray}
where the 12 variational parameters $x_\alpha$, $y_\beta$, $z_\gamma$ and $w_\delta$ are {\it real numbers}. By construction, this PEPS ansatz bears the correct symmetry structure of the two-component state. Upon optimization, a surprisingly good overlap with the state $|\Psi^{+}_{0}\big> + |\Psi^{-}_{\pi}\big>$ is obtained on a 16-site torus -- see Table~\ref{tab:overlap} -- considering the fact that the bond dimension is still quite small (restricting the amount of entanglement in the wavefunction) and the number of real parameters limited to only 11 in that case (one such parameter can be set to 1). Also, the weights of the $Q=0$ and $Q=(\pi,\pi)$ components of the optimized PEPS are quite close to those of the time evolved state, as shown in Fig.~\ref{fig:finiteQ}. Interestingly, this form of PEPS ansatz, by introducing appropriate time-dependent phases into the variational parameters $x_\alpha$, $y_\beta$, $z_\gamma$ and $w_\delta$ provides also an accurate description of the time-dependent state $|\Psi[t_0]\big>\simeq |\phi_0[t_0]\big>$ given by (\ref{eq:psi-adiab}) for all times $t_0\in[0,T_{\rm drive}]$.  

\subsection{Topological properties}

It has been shown that the adiabatically-prepared state in the high-frequency limit~\cite{Mambrini2024} is a topological Abelian CSL, being (close to) the ground-state of a well-studied chiral Heisenberg model~\cite{Poilblanc2017b,Hasik2022}. At finite frequency, however, no such simple static and local Floquet Hamiltonian exists and the prepared quantum state in the stationnary regime acquires the more complex structure given by (\ref{eq:psi-adiab}).
We provide here a summarizing statement that explains how the topological structure/characterization survives as one goes away from the high-frequency limit. For more technical details readers can refer to Appendix~\ref{app:topo}.

One of the key features of a topologically-ordered phase is its topological ground-state degeneracy. Conveniently, the latter can be inferred from its PEPS representation, even approximate, and in particular from the gauge structure of the local PEPS tensor~\cite{Schuch2013a}. In our case, a local $\mathbb{Z}_2$ gauge symmetry is inherited from the SU$(2)$ fusion rules of the virtual space $\nu^{\otimes 4}$ into the physical spin-1/2 space. 
Indeed, one can check explicitly that all eight tensors appearing in (\ref{eq:peps_tensors}) possess such a gauge symmetry (see Appendix~\ref{app:peps}).
To construct different topological PEPS partners on the torus, it is possible to insert $\mathbb{Z}_2$-fluxes in one of the two "holes" of the torus or in both, providing 4 PEPS ans\"atze containing 0, 1 or 2 $\mathbb{Z}_2$ fluxes (there are two choices to insert a single flux on the torus). By investigating the corresponding $4\times 4$ overlap matrix (see Appendix~\ref{app:topo}), one can argue that the dimension of the Hilbert space spanned by these four states is only two, instead of four, suggesting a two-fold topological degeneracy as in the infinite-frequency limit. 

In addition to the initial PEPS ansatz $|\Psi_{\rm PEPS}^{0}\big>$ providing a faithful description of the prepared CSL state, we can then (using $\mathbb{Z}_2$-flux insertions and a linear combination of the four resulting states) construct its orthogonal topological pattern $|\Psi_{\rm PEPS}^{1}\big>$ as shown in Appendix~\ref{app:topo}.  
In close similarity with the limit of the static Floquet Hamiltonian, these two variational states have significant overlaps with the $\alpha=0$ and $\alpha=1$ Floquet eigenstates, as seen in Table~\ref{tab:overlap}, hence providing the identification of the two topological partners constituting the Floquet GS manifold.

\section{Discussions and conclusions} %
\label{sec:conclusion}

In this work we have considered Floquet engineering of a Dynamical Chiral Spin Liquid (DCSL). A special emphasis has been set on the role played by a finite Floquet driving frequency, as well as  on the similarities and differences compared to the high-frequency regime, for which a mapping onto a simple {\it static and local} effective chiral Hamiltonian is possible. In contrast, a finite frequency of the drive (turned on adiabatically) leads to a more complex type of stationary regime with an intrinsic periodic micromotion showing Rabi oscillations caused by interferences between four well-identified components. This DCSL phase is continuously connected to the high-frequency limit, the weights of three of the components vanishing in that limit to recover the static CSL. We also show that, under an adiabatic connection of the drive, the system wavefunction matches with high fidelity the Floquet "ground state" at the edge of the Floquet band (provided the latter does not occupy the full FBZ). The first Floquet excited state is identified as a topological partner, suggesting our DCSL bears $\mathbb{Z}_2$ topological order, in close similarity with its static analog. An accurate description in terms of a (dynamical) PEPS ansatz is provided.

Our calculations have been performed on a finite periodic cluster (torus). It should be mentioned that, keeping the frequency fixed, the thermodynamic limit cannot be taken right-away. Indeed, from (\ref{bandwidth}) we see that a chaotic regime is expected when $N\gtrsim \frac{4}{a}(\omega/J)^2$. One option could be to take some hybrid limit, increasing simultaneously frequency and system size while keeping the ratio $N/\omega^2$ fixed so that the gap $\Delta_N$ normalized by the extension $\omega$ of the FBZ will approach a (non-vanishing) finite limit. It is not clear nevertheless whether this particular limit flows towards the simple static case or rather keeps some of the peculiar features of the DCSL phase. 

In any case, considering cold-atom simulators, one could exploit the finite size of the system and the existence of two finite size gaps, the Floquet gap $\Delta_N(\omega)$ and the level spacing $\delta=\epsilon_1-\epsilon_0$ between the GS and the 1st excited state of the Floquet spectrum, to perform adiabatic quantum state preparation. Note that typically $\Delta_N(\omega)\gg\delta$ so that the adiabaticity is controlled by $\delta$.


As a possible perspective, it would be interesting to explore the conditions under which such driven (non-integrable) spin-liquid models can enter an anomalous topological phase, similar to the phenomenon that was identified for non-interacting fermions~\cite{Rudner2013} and for the driven Kitaev model~\cite{anomalous_Floquet_Kitaev_model}. More generally, it would be intriguing to identify universal and measurable signatures of anomalous topological phases in the  context of correlated quantum matter~\cite{Ryu_Floquet_interacting,Interacting_Floquet_invariants,Interacting_Floquet,Lucila_Streda_Floquet}, which are also accessible in spin systems.

\appendix

\section{Spectrum of the Floquet Hamiltonian}
\label{app:Floquet}

In the case of a microscopic time-dependent Hamiltonian $H(t)$, the unitary time evolution operator involves time-ordering, see~(\ref{eq:Floquet_unit}). In practice, one needs to discretize the time interval by regular time steps $\tau=t/M$. 
Then, a Trotter-Suzuki (TS) decomposition~\cite{SUZUKI1990,Trotter1959}  of the unitary time evolution can be performed in term of elementary gates,
\begin{equation}
    U(t)\simeq\exp{(-iH(t_{M})\tau)}\cdots \exp{(-iH(t_{2})\tau)}\exp{(-iH(t_1)\tau)},
    \label{eq:time-order}
\end{equation}
where $t_n=(n-\frac{1}{2})\tau$ and $M\tau = t$. To check the (formal) $\tau\rightarrow 0$ limit is reached with a good accuracy we have used different values of $M$ to check convergence (reached for $N=64$).

\begin{figure}[h]
	\centering
	\includegraphics[width=0.97\columnwidth]{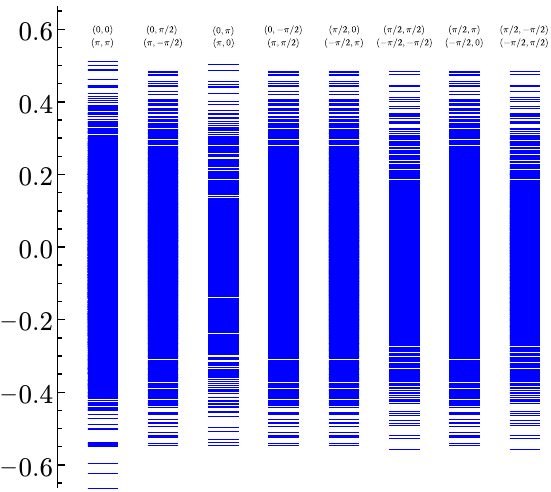}
	\caption{Floquet quasi-energy spectrum for $\omega=4$ with respect to momentum. Energy on the vertical axis is in units of $\pi/T_{\rm drive}=\omega/2$. The different columns are labeled by pairs of momenta $Q$ and $(\pi,\pi)+Q$. 
  }
	\label{fig:spectrum1}
\end{figure}

\begin{figure}
	\centering
	\includegraphics[width=0.97\columnwidth]{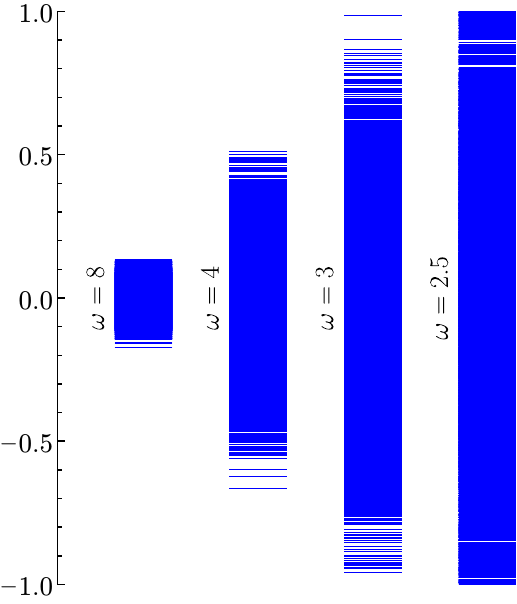}
	\caption{Floquet quasi-energy spectrum for several frequencies. Energies on the vertical axes are in units of $\pi/T_{\rm drive}=\omega/2$. Note that for $\omega=2.5$ the quasi-energy spectrum covers the full Floquet-Brillouin interval $[-\omega/2,\omega/2]$. 
  }
	\label{fig:spectrum2}
\end{figure}

To compute the Floquet Hamiltonian defined by (\ref{eq:Floquet_unit}) one has first to compute and diagonalize the time evolution operator over one period $t=T_{\rm drive}$. The eigenvalues of $U(T_{\rm drive})$ are complex numbers on the unit circle $\{\exp{(-i 2\pi\theta_\alpha)}\}$ from which we can get the quasi-energy spectrum $\{E_\alpha=\frac{2\pi}{T_{\rm drive}}\theta_\alpha\}$.
To reduce the computational cost, some symmetries of $H(t)$ can be used to perform the time-ordered product in each symmetric blocks independently. Taking advantage of the spin-rotation symmetry, the $S_z=0$ basis can be used ($12\, 870$ states for $N=16$ qubits). Within this subspace we have further block-diagonalized the Floquet evolution operator considering lattice translations. The invariance of $H_F$ under the {\it dressed} lattice translation $\tilde T_x$ and $\tilde T_y$ implies to combine the sectors of momentum $Q$ and $(\pi,\pi)+Q$. 
Results for $\omega=4$ are shown in Fig.~\ref{fig:spectrum1} for all (mixed) momentum sectors. 
Comparison of the full spectra (i.e. including all momentum sectors) at different frequencies are shown in Fig.~\ref{fig:spectrum2}. 

\section{Symmetry properties of the Floquet Hamiltonian}
\label{app:FloquetSym}
In this appendix we discuss some useful elementary symmetry properties of $H_F[t_0]$ justifying the Floquet eigenstates form given by Eq.~(\ref{eq:phi}), (\ref{eq:phi2}), (\ref{eq:phi-decomp}) and (\ref{eq:phi-decomp2}).

For an arbitrary time $t_0$ symmetries of $H_F[t_0]$ are fully determined by those of $H(t)$ as defined in Eq.~(\ref{eq:Htotal}). $H(t)$ explicitly breaks translation and $C_4$ point group symmetries, but each terms $H_0$ and $H_{\text{drive}}(t)$ can be decomposed on the irreducible representations of both groups labeled by $Q_{\cal S}$ where $Q$ is the momentum and ${\cal S} = A, B, E$ or $E'$.

$H_0$ obviously belong to the $(0,0)_A$ irreducible representation. $H_{\text{drive}}(t)$ can be split in $H^{+}_{\text{drive}}(t)+H^{-}_{\text{drive}}(t)$ with
\begin{equation}
H^{\pm}_{\text{drive}}(t) = \frac{1}{2} e^{\pm i \omega t} \left ( 
H_x \mp i H_y \right ).
\end{equation}
It is easy to check that $H^{+}_{\text{drive}}(t)$ and $H^{-}_{\text{drive}}(t)$ belong respectively to the $(\pi,\pi)_E$ and $(\pi,\pi)_{E'}$ symmetry sectors.

These remarks have direct consequences on the symmetry structure of the Floquet eigenstates (see Table~\ref{tab:irrepprod}). Starting from the action of the three components of $H(t)$ on a $(0,0)_A$ state allows to identify the stable subspace ${\cal E} = \{ (0,0)_A, (0,0)_B, (\pi,\pi)_E,  (\pi,\pi)_{E'}\}$. On the other hand, the same action on a $(\pi,\pi)_A$ state generates the stable subspace $\tilde{\cal E} = \{  (\pi,\pi)_A, (\pi,\pi)_B, (0,0)_E,  (0,0)_{E'}\}$. As a consequence, any eigenstate of $H_F[t_0]$ involving $(0,0)$ and $(\pi,\pi)$ momenta can either be expressed in ${\cal E}$ (see Eq.~(\ref{eq:phi-decomp})) or $\tilde{{\cal E}}$ (see Eq.~(\ref{eq:phi-decomp2})).

\begin{table}
    \centering
\begin{center}
\begin{tabular}{ l | c c c} \hline
Operator $\rightarrow$ & $H_0$ & $H^{+}_{\text{drive}}(t)$ & $H^{-}_{\text{drive}}(t)$ \\
Symmetry $\rightarrow$ & $(0,0)_A$ & $(\pi,\pi)_E$ & $(\pi,\pi)_{E'}$ \\
$\downarrow$ State  &  &  &  \\\hline
{\color{blue} $(0,0)_A$} & {\color{blue} $(0,0)_A$}  & {\color{blue} $(\pi,\pi)_E$} & {\color{blue} $(\pi,\pi)_{E'}$} \\   
{\color{blue} $(0,0)_B$} & {\color{blue} $(0,0)_B$} & {\color{blue} $(\pi,\pi)_{E'}$} & {\color{blue} $(\pi,\pi)_E$} \\
{\color{red}$(0,0)_E$} & {\color{red}$(0,0)_E$} & {\color{red}$(\pi,\pi)_B$} & {\color{red} $(\pi,\pi)_A$} \\ 
{\color{red}$(0,0)_{E'}$} & {\color{red}$(0,0)_{E'}$} & {\color{red}$(\pi,\pi)_A$} & {\color{red}$(\pi,\pi)_B$} \\ 
{\color{red}$(\pi,\pi)_A$} & {\color{red}$(\pi,\pi)_A$} & {\color{red}$(0,0)_E$} & {\color{red}$(0,0)_{E'}$} \\ 
{\color{red}$(\pi,\pi)_B$} & {\color{red}$(\pi,\pi)_B$} & {\color{red}$(0,0)_{E'}$} & {\color{red}$(0,0)_E$} \\ 
{\color{blue} $(\pi,\pi)_E$} & {\color{blue} $(\pi,\pi)_E$} & {\color{blue} $(0,0)_B$} & {\color{blue} $(0,0)_A$} \\ 
{\color{blue} $(\pi,\pi)_{E'}$} & {\color{blue} $(\pi,\pi)_{E'}$} & {\color{blue} $(0,0)_A$} & {\color{blue} $(0,0)_B$}  \\ \hline   
\end{tabular}
\end{center}
 \caption{\label{tab:irrepprod} Action of the three components of $H(t)$ on $Q=(0,0)$ and $Q=(\pi,\pi)$ states. The set of 8 states splits into two subspaces (in blue and red) that are stable under the action of $H(t)$.}
\end{table}

\section{PEPS tensors}
\label{app:peps}

We provide here the analytic expressions of the twelve $D=3$ tensors used in this work.
A generic tensor on a lattice site of the square lattice is denoted as $T^\sigma (d,r,u,l)$, where $\sigma=\uparrow,\downarrow$ labels the physical spin $S_z=\pm 1/2$ projection at the 
corresponding lattice site and $d,r,u,l$ stand for the labels of the bond indices (in anticlockwise ordering), i.e. $0$, $1$ and $2$ for the $S_z=0$, $1/2$ and $-1/2$ virtual states, respectively.
The PEPS wavefunction is given by
\begin{equation}
    \Psi(\sigma_1,...,\sigma_N)={\rm tTr}(A^{\sigma_1}A^{\sigma_2}...A^{\sigma_N})\, ,
\end{equation}
where $A^{\sigma_i}$ is the tensor at site $i$ and tTr is the tensor-trace which depends on the 2D spatial arrangement of the tensors. 

The tensor elements of the $A_\alpha$ tensors (whose characters under point group operations are displayed on the first line of Table~\ref{tab:sym}) are given by:
\vspace{0.15cm}

\noindent
$
      A_1^\uparrow (0,0,0,1)=1/2 \hskip 1.0cm   A_1^\downarrow  (0,0,0,2)=1/2          \\                
      A_1^\uparrow (0,0,1,0)=1/2 \hskip 1.0cm  A_1^\downarrow  (0,0,2,0)=1/2         \\                
      A_1^\uparrow (0,1,0,0)=1/2 \hskip 1.0cm  A_1^\downarrow  (0,2,0,0)=1/2      \\                                                                       A_1^\uparrow (1,0,0,0)=1/2\hskip 1.0cm   A_1^\downarrow  (2,0,0,0)=1/2                
      \vspace{0.15cm}   \\                       
      A_2^\uparrow(0,1,1,2)=1/(2\sqrt6) \hskip 1.0cm   A_2^\downarrow(0,1,2,2)=-1/(2\sqrt6)\\
      A_2^\uparrow(0,1,2,1)=-1/\sqrt6\hskip 1.0cm    A_2^\downarrow(0,2,1,2)=1/\sqrt6\\
      A_2^\uparrow(0,2,1,1)=1/(2\sqrt6)\hskip 1.0cm   A_2^\downarrow(0,2,2,1)=-1/(2\sqrt6)\\
      A_2^\uparrow(1,0,1,2)=-1/\sqrt6 \hskip 1.0cm   A_2^\downarrow(1,0,2,2)=-1/(2\sqrt6)\\
      A_2^\uparrow(1,0,2,1)=1/(2\sqrt6)\hskip 1.0cm A_2^\downarrow(1,2,0,2)=1/\sqrt6\\
      A_2^\uparrow(1,1,0,2)=1/(2\sqrt6)\hskip 1.0cm    A_2^\downarrow(1,2,2,0)=-1/(2\sqrt6)\\
      A_2^\uparrow(1,1,2,0)=1/(2\sqrt6)\hskip 1.0cm   A_2^\downarrow(2,0,1,2)=-1/(2\sqrt6)\\
      A_2^\uparrow(1,2,0,1)=1/(2\sqrt6)\hskip 1.0cm    A_2^\downarrow(2,0,2,1)=1/\sqrt6\\
      A_2^\uparrow(1,2,1,0)=-1/\sqrt6\hskip 1.0cm   A_2^\downarrow(2,1,0,2)=-1/(2\sqrt6)\\
      A_2^\uparrow(2,0,1,1)=1/(2\sqrt6)\hskip 1.0cm    A_2^\downarrow(2,1,2,0)=1/\sqrt6\\
      A_2^\uparrow(2,1,0,1)=-1/\sqrt6\hskip 1.0cm     A_2^\downarrow(2,2,0,1)=-1/(2\sqrt6)\\
      A_2^\uparrow(2,1,1,0)=1/(2\sqrt6)\hskip 1.0cm   A_2^\downarrow(2,2,1,0)=-1/(2\sqrt6)  
      \vspace{0.15cm} \\
      A_3^\uparrow(1,1,2,0)=i/(2\sqrt2)   \hskip 1.0cm A_3^\downarrow(1,2,2,0)=i/(2\sqrt2)\\ 
       A_3^\uparrow(1,1,0,2)=-i/(2\sqrt2)             \hskip 1.0cm A_3^\downarrow(1,0,2,2)=-i/(2\sqrt2)\\                                
      A_3^\uparrow(1,2,0,1)=i/(2\sqrt2)             \hskip 1.0cm A_3^\downarrow(2,1,0,2)=-i/(2\sqrt2)\\                
      A_3^\uparrow(1,0,2,1)=-i/(2\sqrt2)             \hskip 1.0cm   A_3^\downarrow(2,2,1,0)=-i/(2\sqrt2)\\                
      A_3^\uparrow(2,1,1,0)=-i/(2\sqrt2)             \hskip 1.0cm   A_3^\downarrow(2,2,0,1)=i/(2\sqrt2)\\                
      A_3^\uparrow(2,0,1,1)=i/(2\sqrt2)             \hskip 1.0cm  A_3^\downarrow(2,0,1,2)=i/(2\sqrt2)\\                
      A_3^\uparrow(0,1,1,2)=i/(2\sqrt2)             \hskip 1.0cm   A_3^\downarrow(0,1,2,2)=i/(2\sqrt2)\\                
      A_3^\uparrow(0,2,1,1)=-i/(2\sqrt2)             \hskip 1.0cm  A_3^\downarrow(0,2,2,1)=-i/(2\sqrt2) \\
$

Similarly, the tensor elements of the $B_\alpha$ tensors (whose characters under point group operations are displayed on the second line of Table~\ref{tab:sym}) are given by:
\vspace{0.15cm}

\noindent
$
      B_1^\uparrow (0,0,0,1)=-1/2 \hskip 1.0cm   B_1^\downarrow  (0,0,0,2)=-1/2          \\                
      B_1^\uparrow (0,0,1,0)=1/2 \hskip 1.0cm  B_1^\downarrow  (0,0,2,0)=1/2         \\                
      B_1^\uparrow (0,1,0,0)=-1/2 \hskip 1.0cm  B_1^\downarrow  (0,2,0,0)=-1/2      \\                                                                       B_1^\uparrow (1,0,1,0)=1/2\hskip 1.0cm   B_1^\downarrow  (2,0,0,0)=1/2                
      \vspace{0.15cm}   \\                       
      B_2^\uparrow(0,1,1,2)=1/(2\sqrt6) \hskip 1.0cm   B_2^\downarrow(0,1,2,2)=-1/(2\sqrt6)\\
      B_2^\uparrow(0,1,2,1)=-1/\sqrt6\hskip 1.0cm    B_2^\downarrow(0,2,1,2)=1/\sqrt6\\
      B_2^\uparrow(0,2,1,1)=1/(2\sqrt6)\hskip 1.0cm   B_2^\downarrow(0,2,2,1)=-1/(2\sqrt6)\\
      B_2^\uparrow(1,0,1,2)=1/\sqrt6 \hskip 1.0cm   B_2^\downarrow(1,0,2,2)=-1/(2\sqrt6)\\
      B_2^\uparrow(1,0,2,1)=-1/(2\sqrt6)\hskip 1.0cm B_2^\downarrow(1,2,0,2)=1/\sqrt6\\
      B_2^\uparrow(1,1,0,2)=1/(2\sqrt6)\hskip 1.0cm    B_2^\downarrow(1,2,2,0)=1/(2\sqrt6)\\
      B_2^\uparrow(1,1,2,0)=-1/(2\sqrt6)\hskip 1.0cm   B_2^\downarrow(2,0,1,2)=1/(2\sqrt6)\\
      B_2^\uparrow(1,2,0,1)=1/(2\sqrt6)\hskip 1.0cm    B_2^\downarrow(2,0,2,1)=-1/\sqrt6\\
      B_2^\uparrow(1,2,1,0)=1/\sqrt6\hskip 1.0cm   B_2^\downarrow(2,1,0,2)=-1/(2\sqrt6)\\
      B_2^\uparrow(2,0,1,1)=-1/(2\sqrt6)\hskip 1.0cm    B_2^\downarrow(2,1,2,0)=-1/\sqrt6\\
      B_2^\uparrow(2,1,0,1)=-1/\sqrt6\hskip 1.0cm     B_2^\downarrow(2,2,0,1)=-1/(2\sqrt6)\\
      B_2^\uparrow(2,1,1,0)=-1/(2\sqrt6)\hskip 1.0cm   B_2^\downarrow(2,2,1,0)=1/(2\sqrt6)  
      \vspace{0.15cm} \\
      B_3^\uparrow(1,1,2,0)=i/(2\sqrt2)           \hskip 1.0cm B_3^\downarrow(1,2,2,0)=i/(2\sqrt2)\\ 
       B_3^\uparrow(1,1,0,2)=i/(2\sqrt2)             \hskip 1.0cm B_3^\downarrow(1,0,2,2)=-i/(2\sqrt2)\\                                
      B_3^\uparrow(1,2,0,1)=-i/(2\sqrt2)             \hskip 1.0cm B_3^\downarrow(2,1,0,2)=i/(2\sqrt2)\\                
      B_3^\uparrow(1,0,2,1)=-i/(2\sqrt2)             \hskip 1.0cm   B_3^\downarrow(2,2,1,0)=-i/(2\sqrt2)\\                
      B_3^\uparrow(2,1,1,0)=-i/(2\sqrt2)             \hskip 1.0cm   B_3^\downarrow(2,2,0,1)=-i/(2\sqrt2)\\                
      B_3^\uparrow(2,0,1,1)=i/(2\sqrt2)             \hskip 1.0cm  B_3^\downarrow(2,0,1,2)=i/(2\sqrt2)\\                
      B_3^\uparrow(0,1,1,2)=-i/(2\sqrt2)             \hskip 1.0cm   B_3^\downarrow(0,1,2,2)=-i/(2\sqrt2)\\                
      B_3^\uparrow(0,2,1,1)=i/(2\sqrt2)             \hskip 1.0cm  B_3^\downarrow(0,2,2,1)=i/(2\sqrt2)   \\                
$

We now move to the $E_\alpha$ tensors (whose characters under point group operations are displayed on the third line of Table~\ref{tab:sym}).
The matrix elements of $E_1$ are given by:
\vspace{0.15cm}

\noindent
$
      E_1^\uparrow (0,0,0,1)=i/2 \hskip 1.0cm   E_1^\downarrow  (0,0,0,2)=i/2          \\                
      E_1^\uparrow (0,0,1,0)=-1/2 \hskip 1.0cm  E_1^\downarrow  (0,0,2,0)=-1/2         \\                
      E_1^\uparrow (0,1,0,0)=-i/2 \hskip 1.0cm  E_1^\downarrow  (0,2,0,0)=-i/2      \\                                                                       E_1^\uparrow (1,0,0,0)=1/2\hskip 1.0cm   E_1^\downarrow  (2,0,0,0)=1/2      \\
$

To obtain the remaining $E_2$ and $E_3$ tensors, it is convenient to first define two intermediate $C_{4v}$ "p-wave" $p_a^\sigma$ and $p_b^\sigma$ tensors,
\vspace{0.15cm}

\noindent
$
      p_a^\uparrow(2,1,1,0)=(i/2)/\sqrt6 \hskip 1.0cm   p_a^\downarrow(1,2,2,0)=(-i/2)/\sqrt6\\
      p_a^\uparrow(2,1,0,1)=-1/(2\sqrt6)  \hskip 1.0cm   p_a^\downarrow(1,2,0,2)=1/(2\sqrt6)\\
      p_a^\uparrow(2,0,1,1)=i/\sqrt6     \hskip 1.0cm    p_a^\downarrow(1,0,2,2)=-i/\sqrt6\\
      p_a^\uparrow(1,2,1,0)=(i/2)/\sqrt6     \hskip 1.0cm     p_a^\downarrow(2,1,2,0)=(-i/2)/\sqrt6\\
      p_a^\uparrow(1,2,0,1)=1/\sqrt6            \hskip 1.0cm    p_a^\downarrow(2,1,0,2)=-1/\sqrt6\\
      p_a^\uparrow(1,1,2,0)=-i/\sqrt6          \hskip 1.0cm   p_a^\downarrow(2,2,1,0)=i/\sqrt6\\
      p_a^\uparrow(1,1,0,2)=-1/(2\sqrt6)     \hskip 1.0cm     p_a^\downarrow(2,2,0,1)=1/(2\sqrt6)\\
      p_a^\uparrow(1,0,2,1)=(-i/2)/\sqrt6   \hskip 1.0cm    p_a^\downarrow(2,0,1,2)=(i/2)/\sqrt6\\
      p_a^\uparrow(1,0,1,2)=(-i/2)/\sqrt6 \hskip 1.0cm      p_a^\downarrow(2,0,2,1)=(i/2)/\sqrt6\\
      p_a^\uparrow(0,2,1,1)=1/(2\sqrt6)  \hskip 1.0cm    p_a^\downarrow(0,1,2,2)=-1/(2\sqrt6)\\
      p_a^\uparrow(0,1,2,1)=1/(2\sqrt6) \hskip 1.0cm   p_a^\downarrow(0,2,1,2)=-1/(2\sqrt6)\\ 
      p_a^\uparrow(0,1,1,2)=-1/\sqrt6 \hskip 1.0cm  p_a^\downarrow(0,2,2,1)=1/\sqrt6
      \vspace{0.15cm} \\
      p_b^\uparrow(2,1,1,0)=(i/2)/\sqrt2 \hskip 1.0cm   p_b^\downarrow(1,2,2,0)=(-i/2)/\sqrt2\\
      p_b^\uparrow(2,1,0,1)=1/(2\sqrt2)  \hskip 1.0cm   p_b^\downarrow(1,2,0,2)=-1/(2\sqrt2)\\
      p_b^\uparrow(1,2,1,0)=(-i/2)/\sqrt2     \hskip 1.0cm     p_b^\downarrow(2,1,2,0)=(i/2)/\sqrt2\\
      p_b^\uparrow(1,1,0,2)=-1/(2\sqrt2)     \hskip 1.0cm     p_b^\downarrow(2,2,0,1)=1/(2\sqrt2)\\
      p_b^\uparrow(1,0,2,1)=(-i/2)/\sqrt2   \hskip 1.0cm    p_b^\downarrow(2,0,1,2)=(i/2)/\sqrt2\\
      p_b^\uparrow(1,0,1,2)=(i/2)/\sqrt2 \hskip 1.0cm      p_b^\downarrow(2,0,2,1)=(-i/2)/\sqrt2\\
      p_b^\uparrow(0,2,1,1)=1/(2\sqrt2)  \hskip 1.0cm    p_b^\downarrow(0,1,2,2)=-1/(2\sqrt2)\\
      p_b^\uparrow(0,1,2,1)=-1/(2\sqrt2) \hskip 1.0cm   p_b^\downarrow(0,2,1,2)=1/(2\sqrt2)\\ 
$

 To obtain the correct characters of the $E_\alpha$ tensors (Table~\ref{tab:sym}) a simple unitary transformation is performed:
\begin{eqnarray}
          E_2^\sigma &=& \frac{1}{2} p_a^\sigma - \frac{\sqrt3}{2} p_b^\sigma \\ \nonumber
      E_3^\sigma &=& -i (\frac{\sqrt3}{2} p_a^\sigma + \frac{1}{2} p_b^\sigma) \, .
\end{eqnarray}
    
Finally, the $E'_\alpha$ tensors are just given by $E'_\alpha=(E_\alpha)^*$.

\section{On topological degeneracy}
\label{app:topo}

The PEPS representation of the adiabatically-prepared quantum state, even approximate, provides a very efficient tool to address the topological properties of the phase. The latter is directly connected to the gauge structure of the PEPS. A local $\mathbb{Z}_2$ gauge symmetry is inherited from the SU$(2)$ fusion rules of the virtual space $\nu^{\otimes 4}$ into the physical spin-1/2 space~\cite{Schuch2013a}. In other words, only an {\it odd number} (i.e. one or three) of virtual spin-1/2 can fuse into a physical spin-1/2, this fusion rule being encoded locally in the site tensor. To construct different topological PEPS partners on the torus, it is possible to insert $\mathbb{Z}_2$-fluxes in one of the two "holes" of the torus or in both, providing 4 PEPS ans\"atze. In practice, the flux is inserted by applying a close string of operators $\big(\begin{smallmatrix}
  1 & 0\\
  0 & -1
\end{smallmatrix}\big)$ on the $x$ ($y$) bonds along the $y$ ($x$) direction acting on the virtual space $\nu = 0 \oplus 1/2$. In other words, the string "counts" the number $n_{1/2}$ of virtual spin-1/2 along the close loop and inserts a factor $(-1)^{n_{1/2}}$ to every configuration.

From the optimized PEPS $|\Psi_{\rm PEPS}^0\big>$ one can construct the four (normalized) states $|\Psi_{\rm PEPS}(\alpha_x,\alpha_y)\big>$, $\alpha_x,\alpha_y=0,1$ corresponding to the number of flux quantum inserted along the $y$ and/or $x$ directions. 
Diagonalizing the corresponding $4\times 4$ overlap matrix ${\cal O}_{\alpha\beta}$ enables, in principle, to give the dimension of the space spanned by these states, i.e. the topological degeneracy, but the procedure is subject to strong finite size effects. Nevertheless the limit $1/\omega\rightarrow 0$ of a static Floquet Hamiltonian (\ref{eq:FloquetHam}) is simpler to analyze. In that case, the two leading eigenvalues of $\cal O$ are well separated from the two others (even on a $4\times 4$ torus), suggesting a two-fold topological degeneracy. In fact, the state $|\Psi_{\rm PEPS}(0,0)\big>$ is fully symmetric (only the $A$ component in (\ref{eq:psi-adiab}) and (\ref{eq:peps_tensors}) survives in that limit) and, then, one can construct an orthogonal state $|\Psi_{\rm PEPS}(0,1)\big>- |\Psi_{\rm PEPS}(1,0)\big>$ (belonging to the $B$ IRREP, in agreement with Ref.~\cite{Nielsen2013}) defining the second topological sector. 

At finite frequency, on a torus geometry, the two topological partners should belong to the same symmetry sector since the $A$ and $B$ sectors are coupled. The diagonalization of the $4\times 4$ overlap matrix gives a clear hierarchy of eigenvalues, e.g. for $\omega=4$ we find $\sim 2.5, 1.1, 0.27$ and $0.07$, suggesting a two-dimensional Hilbert space spanned by the 4 states (for larger systems). Then, for a more refined analysis, it is sufficient to consider the space spanned by only the 3 states  $|\Psi_{\rm PEPS}(0,0)\big>=|\Psi_{\rm PEPS}^0\big>$, $|\Psi_{\rm PEPS}(1,0)\big>$ and $|\Psi_{\rm PEPS}(0,1)\big>$.
Let us focus on $\omega=4$ as an example; we find an overlap matrix
\begin{equation}
{\cal O}\simeq \begin{pmatrix}
    1 & 0.770 & 0.788 \\
    0.770 & 1 & 0.386 \\
    0.788 & 0.386 &  1 \\
\end{pmatrix} \, , \nonumber
\end{equation}
whose eigenvalues are (approximately) $2.31, 0.62$ and $0.07$, corresponding to the eigenvectors (displayed in columns),
\begin{equation}
{\cal V}\simeq \begin{pmatrix}
    1.2 & 0.015 & -1.6  \\
    0.99 &-1.03 & 0.95 \\
    1 & 1 &  1 \\
\end{pmatrix} \, . \nonumber
\end{equation}
This suggests that we can choose the (approximately orthogonal) two states $|\Psi_{\rm PEPS}^0\big>=|\Psi_{\rm PEPS}(0,0)\big>$ and  $|\Psi_{\rm PEPS}^1\big>|\propto|\Psi_{\rm PEPS}(1,0)\big> - |\Psi_{\rm PEPS}(0,1)\big>$ to (approximately) span 
the topological subspace, in close similarity with the limit of the static Floquet Hamiltonian. As seen in Table~\ref{tab:overlap}, these two variational states have significant overlaps with the 
$\alpha=0$ and $\alpha=1$ Floquet eigenstates, respectively, believed to be the two topological partners of the Floquet GS manifold.

\vspace{1truecm}
\par\noindent\emph{\textbf{Acknowledgments ---}} %
D.P. acknowledges support from the TNTOP ANR-18-CE30-0026-01 grant awarded by the French Research Council. This work was granted access to the HPC resources of CALMIP center under allocation 2024-P1231. N.~G. is supported by the ERC grant LATIS and the EOS project CHEQS.

\FloatBarrier



\bibliography{bibliography}


\end{document}